\documentclass[journal,11pt,onecolumn]{IEEEtran}

\newcommand\tab[1][1cm]{\hspace*{#1}} 
\usepackage{graphics}
\usepackage{graphicx}
\usepackage{epsfig}
\usepackage{mwe}
\usepackage{subfigure}
\usepackage{float}
\usepackage[justification=centering]{caption}
\usepackage[tbtags]{amsmath}
\usepackage{etoolbox}
\usepackage{bigints}
\usepackage[utf8]{inputenc}
\graphicspath{ {images/} }
\usepackage{mathtools, cuted}
\usepackage{amssymb}
\usepackage{amsthm}
\usepackage{lineno}

\usepackage{booktabs}

\begin{document}
\title{Outage Analysis for SWIPT-Enabled Two-Way Cognitive Cooperative Communications}

\author{Amrita Mukherjee, Tamaghna Acharya,~\IEEEmembership{Member, IEEE}, Muhammad R. A. Khandaker,~\IEEEmembership{Member, IEEE}

\thanks{Amrita Mukherjee and Tamaghna Acharya are with the Department of Electronics and Telecommunication Engineering, Indian institute of Engineering Science and Technology, Shibpur, Howrah, West Bengal, 711103, India. E-mail: amritamukherjee86@yahoo.com; t\textunderscore acharya@iiests.ac.in. \par 
Muhammad R. A. Khandaker is with the Department of Electronic and Electrical Engineering, University College London (UCL), London, U.K. E-mail: m.khandaker@ucl.ac.uk.}
\thanks{ }}

\markboth{ }
{}

\maketitle
\vspace{-1.5cm}
\begin{small}
\section*{Abstract}
In this paper, we study a cooperative cognitive radio network (CCRN) where the secondary user-transmitter (SU-Tx) assists bi-directional communication between a pair of primary users (PUs) following the principle of two-way relaying. In return, it gets access to the spectrum of the PUs to enable its own transmission to SU-receiver (SU-Rx). Further, in order to support sustainable operation of the network, SU-Tx is assumed to harvest energy from the RF signals received from the PUs, using the technique of simultaneous wireless information and power transfer (SWIPT). Assuming a decode-and-forward behaviour and power-splitting based relaying protocol at SU-Tx, closed form expressions for outage probability of PU and SU are obtained. Simulation results validate our analytical results and illustrate spectrum-efficiency and energy-efficiency advantages of the proposed system over one-way relaying.
\end{small}

\begin{small}
\section*{Keywords}
Cooperative cognitive radio network, simultaneous wireless information and power transfer, two-way relaying, decode-and-forward relaying.
\end{small}
\IEEEpeerreviewmaketitle
\vspace{0.5cm}
\section{INTRODUCTION}
\tab Two-way relaying (TWR) \cite{shah2016throughput} is being investigated as a spectrally efficient means for supporting bi-directional communication between a pair of users. In TWR, a single relay node receives the messages from both the users in the first phase using an appropriate multiple access scheme and then, after decoding, compressing or combining them, broadcasts it in the second phase. Finally, the users apply self-interference cancellation technique to retrieve their desired messages from the received broadcast.\\
\tab In a cooperative cognitive radio network (CCRN) \cite{han2009cooperative}, secondary user (SU) is allowed to access the spectrum of primary user (PU) at regular intervals without performing spectrum sensing. Use of CCRN in a TWR system is expected to further improve the spectrum utilization efficiency. In such a system, the SU performs TWR to assist bi-directional communication between a pair of PUs. In return, it secures access to the licensed spectrum of the PUs. In \cite{li2011cognitive}, a spectrum sharing three-phase protocol is proposed based upon decode-and-forward (DF)-based digital network coding. In \cite{li2012spectrum}, a spectrum sharing two-phase protocol is proposed following the principle of analog network coding (ANC) \cite{ahlswede2000network}. The authors derive closed-form expressions for the outage probabilities for both the primary and secondary systems and identify a spectrum sharing region within which the protocol achieves better PU outage performance than the case of direct transmission without spectrum sharing. Similar study in an underlay model is reported in \cite{zhang2015exact}.\\ 
\tab Nevertheless, energizing the SU node by a constant power supply or recharging or replacing its batteries regularly may not be feasible in many applications. Simultaneous Wireless Information and Power Transfer (SWIPT) \cite{zhou2013wireless} is fast emerging as a promising technique to support sustainable network operation in presence of the said challenges. Following this technique, the SU node exploits the RF signal received from the PU to harvest necessary energy to enable relaying of the message of the later as well as transmission of its own message to another SU. Outage performance analysis is carried out in a SWIPT-enabled CCRN, assuming amplify-and-forward (AF) relaying with power splitting (PS) protocol in \cite{wang2014outage}. 
Outage performance in a SWIPT-enabled CCRN under a Nakagami fading channel is analyzed in \cite{jain2015energy} considering decode-and-forward (DF) relaying. The authors in \cite{im2015outage} and \cite{kalamkar2015outage} investigate the performance of a similar SWIPT-enabled CR network in underlay mode.
\\ \tab In SWIPT-enabled two-way CCRN networks, the SU scavenges energy from the RF signals transmitted by both the PUs as well as access PU spectrum in return of its assistance in two-way PU communication. However, to the best of our knowledge, outage performance, spectral-efficiency and energy-efficiency of such networks are not well investigated. A joint power allocation and relay selection scheme is proposed under the constraints of transmit power and interference to the PU \cite{lu2014simultaneous}. In \cite{wang2014wireless}, the authors report closed-form expressions for the outage probability of PU and SU in a SWIPT-enabled two-way CCRN. However, the authors consider AF relaying for the SU- transmitter (SU-Tx) and presence of complex receiver architectures at the PUs to cancel the interference effects caused by SU transmission. In this paper, we consider a similar model with DF relaying. Our additional contributions may be summarized as follows. 
\begin{itemize}
\item We derive exact closed-form expressions for the outage probability of both PU and SU sytems, avoiding the need for sophisticated signal processing operation at PUs. The analytical results are also verified through Monte Carlo simulations. 
\item The impacts of various system parameters, such as power splitting factor for SWIPT and relay location, on the system outage performance are analyzed. 
\item Results of performance comparison between our protocol and similar model with one-way relaying in \cite{jain2015energy} is reported to gain insights on the possible advantages of the proposed system in terms of spectrum-efficiency (SE) and energy-efficiency (EE).\\
\end{itemize}
\tab The rest of this paper is organized as follows. The signal and system modelling is proposed in Section II. Section III derives closed-form analytical expressions for PU and SU outage probability. Section IV then provides the numerical and simulation results and their discussions, while Section V concludes the paper.\\ 
\tab \textbf{Notations: }$Pr\lbrace \cdot \rbrace$ is the outage probability, $f_{X}(x)$ is probability density functions (PDF) of the exponentially distributed random variable X having parameter $\lambda$ denoted by $X\sim \text{exp}(\lambda)$ and $\mathcal{K}_{1}(\cdot)$ represents the modified Bessel function of second kind of order 1 \cite{gradshteyn2007table}.

\vspace{0.5cm}
\section{SYSTEM AND SIGNAL MODELLING}
\tab We consider a CCRN consisting of two primary users PU$_1$ and PU$_2$ and two secondary users SU$_1$ and SU$_2$ as depicted in Fig. 1a. The PUs intend to exchange information with each other. Since the distance between them is beyond the range of effective transmission, they will require some cooperation from neighbouring nodes to forward their data aiming to achieve a target rate of $R_p$ at the PU receivers. SU$_1$ agrees to act as the relay to assist the primary transmission while simultaneously transferring its own data to SU$_2$ to meet a target rate of $R_s$. Furthermore, each node is assumed to be equipped with a single antenna. 

\vspace*{-4mm}
\begin{figure}[htp]
\centering
\includegraphics[width=0.8\linewidth]{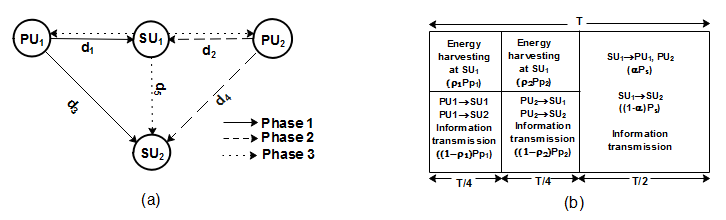}
\caption{(a) System model, (b) Transmission frame structure}
\label{f1}
\end{figure}

\tab We assume that PU$_i$, $i=1,2$ uses constant transmit power $P_{p_{i}}$, while no conventional source of energy is provided to SU$_1$. Thus, it requires to harvest energy from the received signals. Here all channels are assumed to experience independent and identically distributed (i.i.d) Rayleigh fading \cite{fang2015distributed}. The channel coefficients of the links PU$_1\rightarrow$SU$_1$, PU$_2\rightarrow$SU$_1$, PU$_1\rightarrow$SU$_2$, PU$_2\rightarrow$SU$_2$, SU$_1\rightarrow$SU$_2$, SU$_1\rightarrow$PU$_1$ and SU$_1\rightarrow$PU$_2$ are recorded by $h_1$, $h_2$, $h_3$, $h_4$, $h_5$, $h_6$ and $h_7$ respectively with $h_i\sim\mathcal{CN}(0,\Omega_i)$. We have also denoted $ X_i = \vert h_i \vert ^2$. The distances between the users PU$_1-$SU$_1$, PU$_2-$SU$_1$, PU$_1-$SU$_2$, PU$_2-$SU$_2$ and SU$_1-$SU$_2$ are given by $d_1$, $d_2$, $d_3$, $d_4$ and $d_5$ respectively with $m$ as the path-loss exponent. The additive white Gaussian noise (AWGN) at the respective receivers is denoted by $ n_{j}\sim\mathcal{CN}(0,\sigma_j^2)\hspace{1mm}\forall \hspace{1mm} j\in \lbrace{PU_1, PU_2, SU_1, SU_2}\rbrace $ and $ n_{c}\sim\mathcal{CN}(0,\sigma_c^2)$ is the sampled AWGN due to RF to baseband signal conversion. \\
\tab Fig. 1b shows the transmission frame for the two-way relaying protocol which consists of three phases. In phases 1 and 2, PU$_1$ and PU$_2$ transmit their information $X_{p_1}$ and $X_{p_2}$ respectively, to SU$_1$. However, SU$_2$ can also receive the primary signals because of the broadcast nature of the wireless medium. After harvesting energy from part of the received signals in the two phases separately, SU$_1$ broadcasts a network-coded \cite{ahlswede2000network} primary signal $X_{p_1} \oplus X_{p_2}$ (bitwise XOR) superposed with the secondary signal $ X_s $ in the third phase, using all of the harvested energy. Once PU$_1$, PU$_2$ and SU$_2$ have received the broadcasted signal from SU$_1$, they can decode the desired information from the mixed signal based on their own and prior information received in phases 1 and 2, as in \cite{jain2015energy}. \\ 
\tab As illustrated in Fig. 1b, the whole process takes place in three phases. In phase 1, the signal received by SU$_1$ and SU$_2$ can be expressed as
\begin{align} \label{e1}
Y_{SU_{1}}^{(1)} = \sqrt{\frac{P_{p_{1}}}{d_1^m}}h_1 X_{p_1} + n_{SU_1} ,
\end{align}
\begin{align} \label{e2}
Y_{SU_{2}}^{(1)} = \sqrt{\frac{P_{p_{1}}}{d_3^m}}h_3 X_{p_1} + n_{SU_2} .
\end{align}
\\
\tab Based on the power-splitting method in \cite{zhou2013wireless}, a part of the the received information in SU$_1$ is used for energy harvesting and the harvested energy in this phase is given by
\begin{align} \label{e3}
E_1 = \Bigg(\frac{\eta \rho_1 P_{p_{1}}}{d_1^m}\vert h_1 \vert ^2\Bigg)\frac{T}{4}
\end{align}
where $0\leq \eta\leq 1$ represents the energy conversion efficiency and $0<\rho_1<1$ is the portion of information split for energy harvesting in phase 1 and is referred as power splitting factor in phase 1 in the subsequent discussion.
The signal received at information receiver of SU$_1$ is given by
\begin{align} \label{e4}
\hspace{-2mm}\sqrt{1\hspace{-1mm}-\hspace{-1mm}\rho_1}Y_{SU_{1}}^{(1)}\hspace{-1mm} = \hspace{-1mm}\underbrace{\sqrt{\frac{(1\hspace{-1mm}-\hspace{-1mm}\rho_1) P_{p_{1}}}{d_1^m}}h_1 X_{p_1}}_{\text{required signal}}\hspace{-1mm} + \hspace{-1mm}\underbrace{\sqrt{1\hspace{-1mm}-\hspace{-1mm}\rho_1}n_{SU_1}\hspace{-1mm} + \hspace{-1mm} n_c}_{\text{noise}} .
\end{align}
From (\ref{e4}), the total AWGN variance at SU$_1$ is given by $\sigma^2 = (1-\rho_1)\sigma_{SU_1}^2 + \sigma_c^2$.
Thus, in phase 1, the rate achievable at SU$_1$ will be
\begin{align} \label{e5}
R_{SU_{1}}^{(1)} = \frac{1}{4}\log_2 \Bigg(1+\frac{(1-\rho_1) P_{p_{1}}}{d_1^m \sigma^2}\vert h_1 \vert ^2 \Bigg)
\end{align}
and at SU$_2$ will be 
\begin{align} \label{e6}
R_{SU_{2}}^{(1)} = \frac{1}{4}\log_2 \Bigg(1+\frac{P_{p_{1}}}{d_3^m \sigma_{SU_2}^2}\vert h_3 \vert ^2 \Bigg) .
\end{align}
Similarly, in phase 2, the signal received by SU$_1$ and SU$_2$ can be formulated as
\begin{align} \label{e7}
Y_{SU_{1}}^{(2)} = \sqrt{\frac{P_{p_{2}}}{d_2^m}}h_2 X_{p_2} + n_{SU_1} ,
\end{align}
\begin{align} \label{e8}
Y_{SU_{2}}^{(2)} = \sqrt{\frac{P_{p_{2}}}{d_4^m}}h_4 X_{p_2} + n_{SU_2} .
\end{align}
The harvested energy in this phase is thus given by
\begin{align} \label{e9}
E_2 = \Bigg(\frac{\eta \rho_2 P_{p_{2}}}{d_2^m}\vert h_2 \vert ^2\Bigg)\frac{T}{4}
\end{align}
where $0<\rho_2<1$ is the power splitting factor in phase 2.
Thus, in phase 2, the rate achievable at SU$_1$ will be
\begin{align} \label{e5}
R_{SU_{1}}^{(2)} = \frac{1}{4}\log_2 \Bigg(1+\frac{(1-\rho_2) P_{p_{2}}}{d_2^m \sigma^2}\vert h_2 \vert ^2 \Bigg)
\end{align}
and at SU$_2$ will be 
\begin{align} \label{e6}
R_{SU_{2}}^{(2)} = \frac{1}{4}\log_2 \Bigg(1+\frac{P_{p_{2}}}{d_4^m \sigma_{SU_2}^2}\vert h_4 \vert ^2 \Bigg) .
\end{align}
At the end of phase 2, the total harvested power at SU$_1$ (transmission power of SU$_1$ in phase 3) is given by 
\begin{align} \label{e12}
P = \frac{E_1+E_2}{\frac{T}{2}}
= \frac{\eta}{2} \big( a\vert h_1 \vert ^2 + b\vert h_2 \vert ^2 \big)
\end{align}
where $a \overset{\Delta}{=} \frac{\rho_1 P_{p_{1}}}{d_1^m}$ and $b \overset{\Delta}{=} \frac{\rho_2 P_{p_{2}}}{d_2^m}$. \\
\tab Following the DF strategy, SU$_1$ uses a fraction $\alpha$ ($0<\alpha<1$) of its total transmit power $P$ to relay the network-coded primary information and uses the remaining power to transmit an independent message $X_s$ to SU$_2$. We term $\alpha$ as a power-allocation factor throughout this paper.
Therefore, after cancellation of self-interference, the signal received at PU$_1$ and PU$_2$ in phase 3 can be denoted as
\begin{align} \label{e13}
Y_{PU_{1}}^{(3)} = \underbrace{\sqrt{\frac{\alpha P}{d_1^m}}h_6 X_{p_2}}_{\text{required signal}} + \underbrace{\sqrt{\frac{(1-\alpha)P}{d_1^m}}h_6 X_s}_{\text{interference}} + \underbrace{n_{PU_1}}_{\text{noise}} ,
\end{align}
\begin{align} \label{e14}
Y_{PU_{2}}^{(3)} = \underbrace{\sqrt{\frac{\alpha P}{d_2^m}}h_7 X_{p_1}}_{\text{required signal}} + \underbrace{\sqrt{\frac{(1-\alpha)P}{d_2^m}}h_7 X_s}_{\text{interference}} + \underbrace{n_{PU_2}}_{\text{noise}} .
\end{align}
while that received at SU$_2$, assuming both the primary signals were successfully decoded, is given by
\begin{align} \label{e15}
Y_{SU_{2}}^{(3)} = \underbrace{\sqrt{\frac{(1-\alpha)P}{d_5^m}}h_5 X_s}_{\text{required signal}} + \underbrace{n_{SU_2}}_{\text{noise}} .
\end{align}
Hence, the rate achievable at PU$_1$ is given by
\begin{align} \label{e16}
\begin{split}
R_{PU_{1}}^{(3)} &= \frac{1}{2} \log_2 \Bigg(1+\frac{\frac{\alpha P}{d_1^m} \vert h_6 \vert ^2}{\frac{(1-\alpha)P}{d_1^m} \vert h_6 \vert ^2 + \sigma_{PU_1}^2}\Bigg)\\
&= \frac{1}{2} \log_2 \Bigg(1+\frac{a^\prime \lbrace a\vert h_1 \vert ^2 + b\vert h_2 \vert ^2\rbrace \vert h_6 \vert ^2}{b^\prime \lbrace a\vert h_1 \vert ^2 + b\vert h_2 \vert ^2\rbrace \vert h_6 \vert ^2 + 1}\Bigg)
\end{split}
\end{align}
where $a^\prime \overset{\Delta}{=} \frac{\alpha \eta}{2 d_1^m \sigma_{PU_1}^2}$ and $b^\prime \overset{\Delta}{=} \frac{(1-\alpha) \eta}{2 d_1^m \sigma_{PU_1}^2}$. Similarly, rate achievable at PU$_2$ is given by
\vspace{-5mm}
\begin{align} \label{e17}
\begin{split}
R_{PU_{2}}^{(3)} 
&= \frac{1}{2} \log_2 \Bigg(1+\frac{a^{\prime\prime} \lbrace a\vert h_1 \vert ^2 + b\vert h_2 \vert ^2\rbrace \vert h_7 \vert ^2}{b^{\prime\prime} \lbrace a\vert h_1 \vert ^2 + b\vert h_2 \vert ^2\rbrace \vert h_7 \vert ^2 + 1}\Bigg)
\end{split}
\end{align}
where $a^{\prime\prime} \overset{\Delta}{=} \frac{\alpha \eta}{2 d_2^m \sigma_{PU_2}^2}$ and $b^{\prime\prime} \overset{\Delta}{=} \frac{(1-\alpha) \eta}{2 d_2^m \sigma_{PU_2}^2}$, while at $SU_2$ by
\begin{align} \label{e18}
R_{SU_{2}}^{(3)} = \frac{1}{2} \log_2 \bigg(1+c \big( a\vert h_1 \vert ^2 + b\vert h_2 \vert ^2 \big) \vert h_5 \vert ^2 \bigg)
\end{align}
where $c \overset{\Delta}{=} \frac{(1-\alpha)\eta}{2 d_5^m \sigma_{SU_2}^2}$.

\vspace{0.5cm}
\section{PERFORMANCE ANALYSIS}
\tab In this section, the exact closed-form expressions of the outage probability of PUs and SU are derived for the above two-way energy harvesting DF relaying protocol.
\subsection{\textit{Outage probability of Primary System}}
\tab An outage is declared for the primary system if any of the links in the three phases, i.e., PU$_1\rightarrow$SU$_1$, PU$_2\rightarrow$SU$_1$, SU$_1\rightarrow$PU$_1$ or SU$_1\rightarrow$PU$_2$ fail to achieve the target rate $R_p$ that is required to decode the primary information. Therefore,
\begin{equation} \tag{19}\label{e19}
\begin{split}
P_{out}^{PU} &= 1 - Pr\big\lbrace R_{SU_1}^{(1)}>R_p\big\rbrace Pr\big\lbrace R_{SU_1}^{(2)}>R_p\big\rbrace Pr\big\lbrace min(R_{PU_1}^{(3)},R_{PU_2}^{(3)})>R_p\big\rbrace .
\end{split}
\end{equation}
Since the events in the last probability term are independent of each other, we can write eqn. (\ref{e19}) as
\begin{equation} \tag{20}\label{e20}
\begin{split}
P_{out}^{PU} = 1 - Pr\big\lbrace R_{SU_1}^{(1)}>R_p\big\rbrace Pr\big\lbrace R_{SU_1}^{(2)}>R_p\big\rbrace Pr\big\lbrace R_{PU_1}^{(3)}>R_p\big\rbrace Pr\big\lbrace R_{PU_2}^{(3)}>R_p\big\rbrace .
\end{split}
\end{equation}
\hspace*{3mm}\textit{\textbf{Lemma 1. }}
The outage probability of the primary system is given by
\begin{equation} \tag{21}\label{e21}
\begin{split}
P_{out}^{PU} = \begin{cases}
1 - \bigg(\frac{2 \lambda_1\lambda_2}{a\lambda_2-b\lambda_1}\bigg)^2 \text{exp}\Bigg\lbrace{-{\frac{\lambda_1 \gamma_{p_1} d_1^m \sigma^2}{(1-\rho_1) P_{p_{1}}}}}\Bigg\rbrace \text{exp}\Bigg\lbrace{-{\frac{\lambda_2 \gamma_{p_1} d_2^m \sigma^2}{(1-\rho_2) P_{p_{2}}}}}\Bigg\rbrace 
\Bigg[ \sqrt{\frac{\gamma_{p_2} \lambda_6}{(a^\prime - \gamma_{p_2} b^\prime)\lambda_1}a}\hspace{1mm}\mathcal{K}_1{\Bigg(2 \sqrt{\frac{\gamma_{p_2} \lambda_6 \lambda_1}{a(a^\prime - \gamma_{p_2} b^\prime)}}\Bigg)}\nonumber\\
\hspace{2cm}-\sqrt{\frac{\gamma_{p_2} \lambda_6}{(a^\prime - \gamma_{p_2} b^\prime)\lambda_2}b}\hspace{2mm}\mathcal{K}_1{\Bigg(2 \sqrt{\frac{\gamma_{p_2} \lambda_6 \lambda_2}{b(a^\prime - \gamma_{p_2} b^\prime)}}\Bigg)}\Bigg]\Bigg[ \sqrt{\frac{\gamma_{p_2} \lambda_7}{(a^{\prime\prime} - \gamma_{p_2} b^{\prime\prime})\lambda_1}a}\hspace{1mm}\mathcal{K}_1{\Bigg(2 \sqrt{\frac{\gamma_{p_2} \lambda_7 \lambda_1}{a(a^{\prime\prime} - \gamma_{p_2} b^{\prime\prime})}}\Bigg)}\nonumber\\
\hspace{7.6cm}-\sqrt{\frac{\gamma_{p_2} \lambda_7}{(a^{\prime\prime} - \gamma_{p_2} b^{\prime\prime})\lambda_2}b}\hspace{1mm}\mathcal{K}_1{\Bigg(2 \sqrt{\frac{\gamma_{p_2} \lambda_7 \lambda_2}{b(a^{\prime\prime} - \gamma_{p_2} b^{\prime\prime})}}\Bigg)}\Bigg], \gamma_{p_2}\hspace{-1mm} < \hspace{-1mm}\frac{\alpha}{1-\alpha}\nonumber\\
1, \hspace{2cm}\text{otherwise} .\\
\end{cases}
\end{split}
\end{equation}
where \(\gamma_{p_1}=2^{4 R_p}-1\) and \(\gamma_{p_2}=2^{2 R_p}-1\).\\
\hspace*{3mm}\textit{\textbf{Proof. }}
See Appendix A.
\subsection{\textit{Outage probability of Secondary System}}
\tab Success in SU's transmission is defined as the occurrence of the following joint events:\\
(i) PU$_1$'s transmission is successfully decoded at SU$_1$ and SU$_2$ in Phase 1, \\
(ii) PU$_2$'s transmission is successfully decoded at SU$_1$ and SU$_2$ in Phase 2 and \\
(iii) SU$_1$'s transmission is successfully decoded at SU$_2$ in phase 3. \\
\tab Considering the independence of the above events, the outage of SU for a given target rate $R_s$ is defined as
\begin{equation} \tag{22}\label{e23}
\begin{split}
P_{out}^{SU}\hspace{-1mm} =\hspace{-1mm} 1\hspace{-1mm} -\hspace{-1mm} Pr\big\lbrace R_{SU_1}^{(1)}\hspace{-1mm}>\hspace{-1mm}R_p\big\rbrace Pr\big\lbrace R_{SU_2}^{(1)}\hspace{-1mm}>\hspace{-1mm}R_p\big\rbrace Pr\big\lbrace R_{SU_1}^{(2)}\hspace{-1mm}>\hspace{-1mm}R_p\big\rbrace Pr\big\lbrace R_{SU_2}^{(2)}\hspace{-1mm}>\hspace{-1mm}R_p\big\rbrace Pr\big\lbrace R_{SU_2}^{(3)}\hspace{-1mm}>\hspace{-1mm}R_s\big\rbrace .
\end{split}
\end{equation}
\hspace*{3mm}\textit{\textbf{Lemma 2. }}
The outage probability of the secondary system is given by
\begin{equation} \tag{23}\label{e22}
\begin{split}
P_{out}^{SU} = 1 - \bigg(\frac{2 \lambda_1\lambda_2}{a\lambda_2-b\lambda_1}\bigg) \text{exp}\Bigg\lbrace{-{\frac{\lambda_1 \gamma_{p_1} d_1^m \sigma^2}{(1-\rho_1) P_{p_{1}}}}}\Bigg\rbrace \text{exp}\Bigg\lbrace{-{\frac{\lambda_3 \gamma_{p_1} d_3^m \sigma_{SU_2}^2}{P_{p_{1}}}}}\Bigg\rbrace 
\text{exp}\Bigg\lbrace{-{\frac{\lambda_2 \gamma_{p_1} d_2^m \sigma^2}{(1-\rho_2) P_{p_{2}}}}}\Bigg\rbrace \text{exp}\Bigg\lbrace{-{\frac{\lambda_4 \gamma_{p_1} d_4^m \sigma_{SU_2}^2}{P_{p_{2}}}}}\Bigg\rbrace\\
\times\Bigg[ \sqrt{\frac{\lambda_5 \gamma_s a}{\lambda_1 c}}\hspace{1mm}\mathcal{K}_1{\bigg(2 \sqrt{\frac{\lambda_5 \lambda_1 \gamma_s}{ac}}\bigg)} - \sqrt{\frac{\lambda_5 \gamma_s b}{\lambda_2 c}}\hspace{1mm}\mathcal{K}_1{\bigg(2 \sqrt{\frac{\lambda_5 \lambda_2 \gamma_s}{bc}}\bigg)}\Bigg]
\end{split}
\end{equation}
where \(\gamma_s=2^{2 R_s}-1\). \\
\hspace*{3mm}\textit{\textbf{Proof. }}
See Appendix B.
\subsection{\textit{Spectrum Efficiency (SE) and Energy Efficiency (EE)}}
\tab Results of the outage analysis also help us in evaluating average spectrum-efficiency ($\eta_{SE}$) (or throughput) and average energy-efficiency ($\eta_{EE}$) of the proposed system and compare them to that of similar CCRN with one-way communication. We define them as follows : 
\begin{equation} \tag{24}\label{e24}
\begin{split}
\eta_{SE} &= \text{Average SE of PU(s)}+\text{Average SE of SU}\\
&= 2\frac{T}{2T}R_p (1\hspace{-1mm}-\hspace{-1mm}P_{out}^{PU})+\frac{T}{2T} R_s (1\hspace{-1mm}-\hspace{-1mm}P_{out}^{SU})
\end{split}
\end{equation}
\begin{equation} \tag{25}\label{e25}
\hspace*{-2.3cm}\eta_{EE} = \frac{\eta_{SE}}{\text{Transmit power of PU(s)}} .
\end{equation}
Evaluation of system performance using these metrics will be discussed in the next section.

\vspace{0.5cm}
\section{RESULTS AND DISCUSSIONS}
\tab In this section, simulation results are presented and compared with the analytical results derived in (\ref{e21}) and (\ref{e22}). In addition, variation of PU and SU outage probabilities with various system parameters is also illustrated. Finally, results of SE and EE of the proposed system are also presented. The relative positions of the users are shown in Fig. 1a. Until otherwise specified, the necessary simulation parameters are stated as in Table 1.

\begin{table}[h]
\caption{List of necessary simulation parameters}
\centering
\begin{tabular}{c c}
\hline
Name &Value\\
\hline\hline
Distance between PU$_1$ and PU$_2$, $L$ & $2$ m\\ 
\hline
Noise variance, $\sigma_{PU_1}^2=\sigma_{PU_2}^2=\sigma_{SU_2}^2=\sigma^2$ & $0$ dB\\
\hline
Normalized transmit power, \\$\frac{P_{p_1}}{\sigma^2}=\frac{P_{p_2}}{\sigma^2}$ & $40$ dB\\
\hline
Energy conversion efficiency, $\eta$ & $0.9$\\ 
\hline
Power allocation factor, $\alpha$ & $0.9$\\ 
\hline
Power splitting factor, $\rho_1=\rho_2$ & $0.5$\\ 
\hline
Distance between SU$_1$ and PU$_1$(PU$_2$), $d_1(d_2)=\frac{L}{2}$, & $1$ m\\ 
\hline
Distance between SU$_1$ and SU$_2$ when $d_1=d_2$, $d_5=\frac{L}{2}$ & $1$ m\\ 
\hline
Distance between SU$_2$ and PU$_1$(PU$_2$), $d_3(d_4)=\frac{L}{\sqrt{2}}$ & $1.414$ m\\ 
\hline
Path loss exponent, $m$ & $3$\\
\hline
Target rate, $R_p$,$R_s$ & $1$ bits/s/Hz\\
\hline
\end{tabular}
\end{table}

\begin{figure*}[b]
\centering
  \begin{minipage}{.55\textwidth}
  \includegraphics[width=\textwidth]{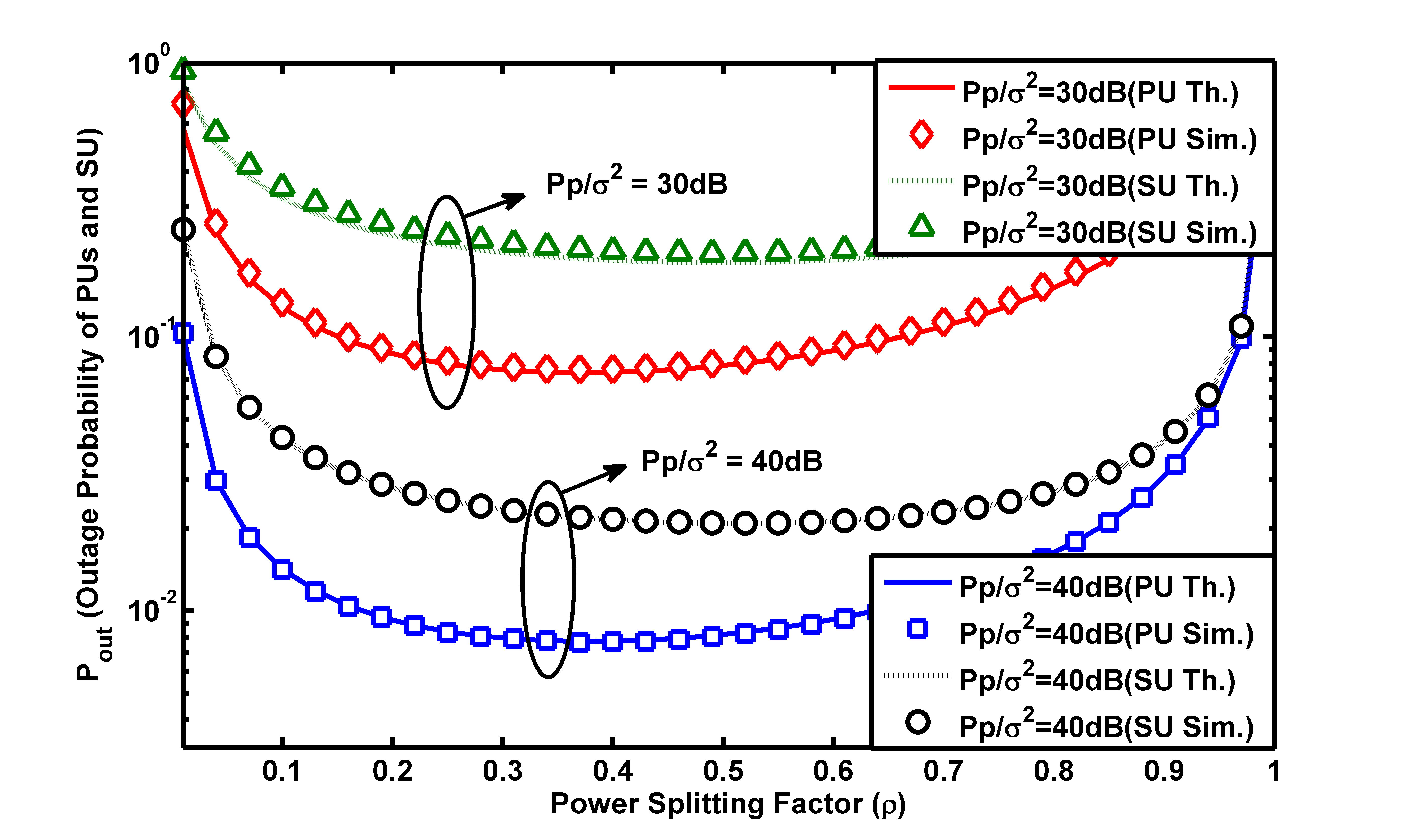}
  \caption{System outage vs Power splitting factor ($\rho$)}
  \label{f3}
  \end{minipage}
\end{figure*}

\tab Fig. \ref{f3} shows the variations of PU and SU outage probabilities with the power splitting factor $\rho$, for two different values of PU's (normalized) transmit power: (i) 30 dB and (ii) 40 dB. An excellent matching between the analytical and simulation results validates the closed form expressions derived in (\ref{e21}) and (\ref{e22}). For both the primary and secondary systems, the result of outage decreases initially when $\rho$ increases from 0 to an optimal value, but it starts increasing beyond this value. From the plots, the optimal value for the primary users is found to be $0.37$. For the primary system, this may be explained as follows. $\rho$ represents the fraction of energy scavenged from the received PU signals in phases 1 and 2. As $\rho$ increases, at first, the amount of transmit power that is used for harvesting energy increases and subsequently less remains for information decoding at SU$_1$. So the probability with which PU signals are successfully decoded at SU$_1$ $\big(\text{i.e. }Pr\big\lbrace R_{SU_1}^{(1)}\hspace{-2mm}>\hspace{-2mm}R_p\big\rbrace$ and $Pr\big\lbrace R_{SU_1}^{(2)}\hspace{-1mm}>\hspace{-1mm}R_p\big\rbrace\big)$ decreases (fig. not shown). At the same time, signals are decoded at the PUs with higher probability, i.e. $Pr\big\lbrace R_{PU_1}^{(3)}\hspace{-1mm}>\hspace{-1mm}R_p\big\rbrace$ and $Pr\big\lbrace R_{PU_2}^{(3)}\hspace{-1mm}>\hspace{-1mm}R_p\big\rbrace$ increases (fig. not shown). This is due to the fact that more power is available for transmission in 
the broadcast phase due to availability of higher energy in SU$_1$. However, it is seen that the increase in successful decoding probabilities in phase 3 overrides the decrease of the same in phases 1 and 2, which results in the initial fall in the outage (refer eqn. (\ref{e19})). Interestingly, as $\rho$ continues to increase beyond the critical value, although the energy harvested also increases, power is increasingly becoming inadequate for information decoding at SU$_1$. Hence, though PU signals are decoded at PU$_1$ and PU$_2$ in phase 3 with higher probability, the decrease in the probability of successful decoding of PU signals at SU$_1$ dominates eqn. (\ref{e19}). Thus a rise in the primary system outage is observed thereafter. A similar line of reasoning would also justify the nature of variation of outage probability of SU. Hence, the secondary outage depends only on the successful decoding probability of PU signals at SU$_1$ in phases 1 and 2 and that at SU$_2$ in phase 3. Initially, as $\rho$ increases, more amount of storage energy is available in SU$_1$ for transmission in phase 3. But the power remaining in SU$_1$ after harvesting is highly insufficient to successfully decode PU signals at SU$_1$ with high probability (fig. not shown). It is observed that the increase in the success probability of decoding SU signal at SU$_2$ $\big(\text{i.e. }Pr\big\lbrace R_{SU_2}^{(3)}\hspace{-1mm}>\hspace{-1mm}R_s\big\rbrace\big)$ dominates the secondary outage (refer eqn. (\ref{e23})). But when $\rho$ exceeds the critical value of $0.52$, the harvested energy becomes so high that the decrease in the successful decoding probability of PU signals at SU$_1$ overrides the increase in the probability with which SU signal is successfully decoded at SU$_2$ in eqn. (\ref{e23}). We, therefore, notice an increase in the secondary system outage subsequently. It can also be observed from the plots that with the increase in the transmit power of each PU from $30$ dB to $40$ dB, there is no change in the optimal power splitting factor but the outage shows a considerable fall by about $89\%$ for both the PUs and SU at the optimal points. This is because with the increase of transmit power, the average value of total harvested energy at SU$_1$ increases for a fixed $\rho$. Thus, its influence on the successful decoding of signals at the destinations is more at higher transmit power and hence the outage improves.

\begin{figure*}[h]
\centering
\begin{minipage}{.5\textwidth}
  \centering                       
  \includegraphics[width=\textwidth]{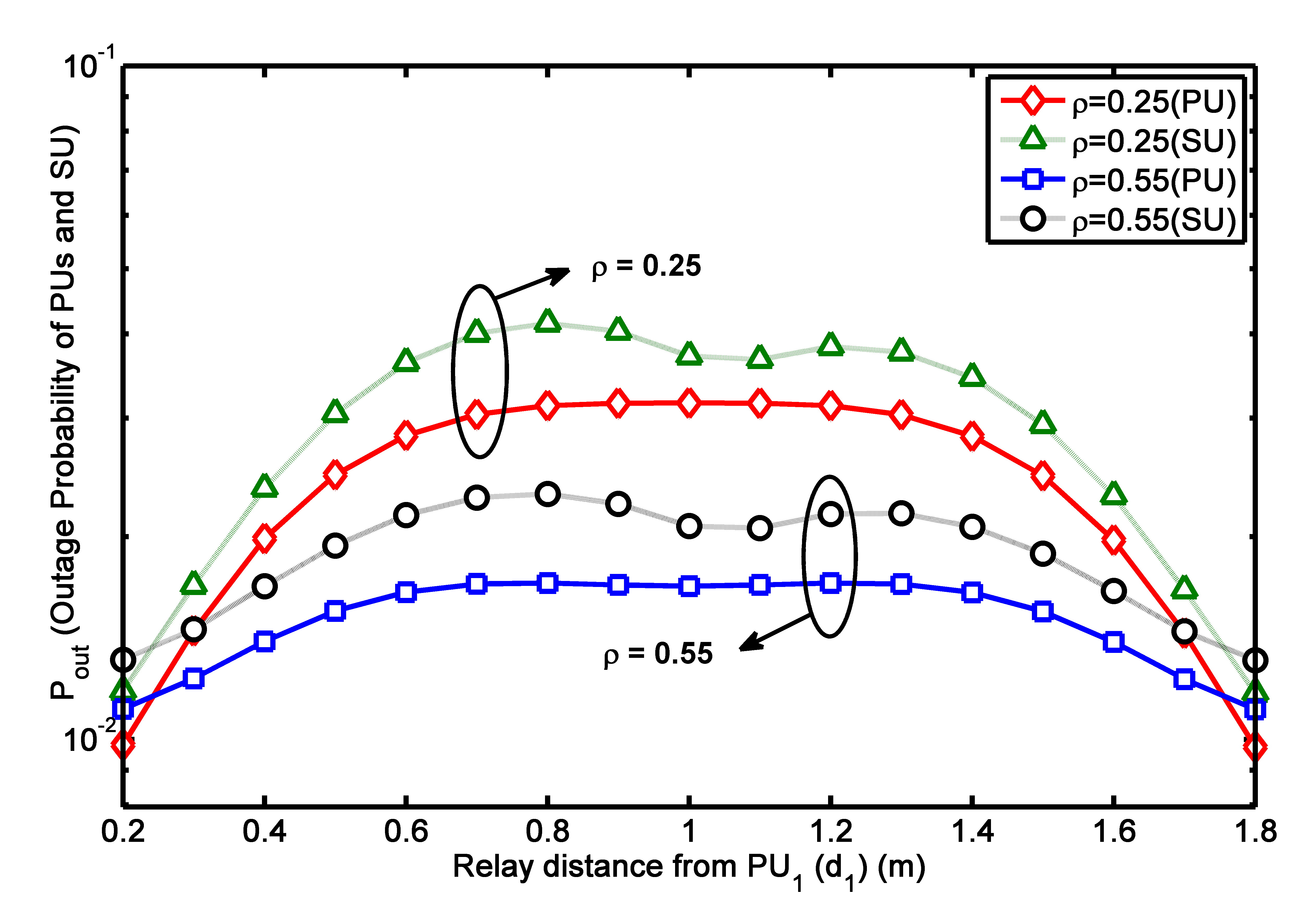}
  \captionof{figure}{System outage vs Relay location from $PU_1$ ($d_1$)}
  \label{f4}
\end{minipage}
\end{figure*}

\tab Fig. \ref{f4} depicts the variation in outage probability of both PUs and SU respectively, with respect to the relay location, for two different values of power splitting factor $\rho$. As SU$_1$ moves away from PU$_1$, both primary and secondary outage increases initially but then it starts decreasing. For the primary system, this can be explained as follows. When distance between SU$_1$ and PU$_1$ increases, though the channel condition between PU$_1-$SU$_1$ worsens, PU$_2-$SU$_1$ link gets better. It is also noted (fig. not shown) that as SU$_1$ moves away from PU$_1$, the success probability of decoding both the PUs' signals at SU$_1$ in phases 1 and 2 $\big(\text{i.e. }Pr\big\lbrace R_{SU_1}^{(1)}\hspace{-2mm}>\hspace{-1mm}R_p\big\rbrace Pr\big\lbrace R_{SU_1}^{(2)}\hspace{-2mm}>\hspace{-1mm}R_p\big\rbrace\big)$ improves, though the average value of the total energy harvested at SU$_1$ from both PU signals decreases. Since the harvested energy is the driving force for the broadcast phase $\big(\text{i.e. }Pr\big\lbrace R_{PU_1}^{(3)}\hspace{-1mm}>\hspace{-1mm}R_p\big\rbrace Pr\big\lbrace R_{PU_2}^{(3)}\hspace{-1mm}>\hspace{-1mm}R_p\big\rbrace\big)$, signals are decoded at the respective PUs with lower probability. This decrease, however, dominates eqn. (\ref{e19}), and hence the outage of the primary system rises. When SU$_1$ moves beyond the critical distance of $1$m from PU$_1$ and hence closer to PU$_2$, the stored energy in SU$_1$ increases due to the scope of harvesting energy at SU$_1$ from both PU$_1$ and PU$_2$. As a result, probability of successful decoding of PU signals in phase 3 increases thereafter and so the outage of the primary system falls. On the other hand, for the secondary system, considering SU$_2$ fixed, as SU$_1$ moves in between the PUs, the distance between SU$_1$ and SU$_2$ changes with minimum being at the centre position. Therefore, as SU$_1$ moves away from PU$_1$, SU$_1-$SU$_2$ link gets better, but then it deteriorates as SU$_1$'s distance from PU$_1$ exceeds $1$ m. Initially, the decoding probabilities of both the PUs' signals at SU$_1$ increases with increasing distance between PU$_1$ and SU$_1$. But since the average value of the total harvested energy at SU$_1$ controls eqn. (\ref{e23}), the successful probability of decoding SU signal at SU$_2$ initially decreases with decreasing harvested energy, resulting in an initial rise in the outage. Afterwards, the outage decreases beyond $1$m with increasing harvested energy. Interestingly, a slight fall in outage is observed at a distance of $1$ m. This can be possibly explained from the fact that though the harvested power is minimum at the centre, the channel condition between SU$_1-$SU$_2$ is the best. Hence the SU outage undergoes a slight improvement at this point, as SU$_1-$SU$_2$ channel condition acts as the dominating factor in eqn. (\ref{e23}). From the plots we can also observe that when $\rho$ increases from $0.25$ to $0.55$, the worst-case outage performance improves by $47\%$ for the PUs and by $44\%$ for the SU. This is because with the increase of $\rho$, harvested energy increases and hence the outage performance gets better.

\begin{figure*}[h]
\centering
\begin{minipage}{.55\textwidth}
  \centering
  \includegraphics[width=\textwidth]{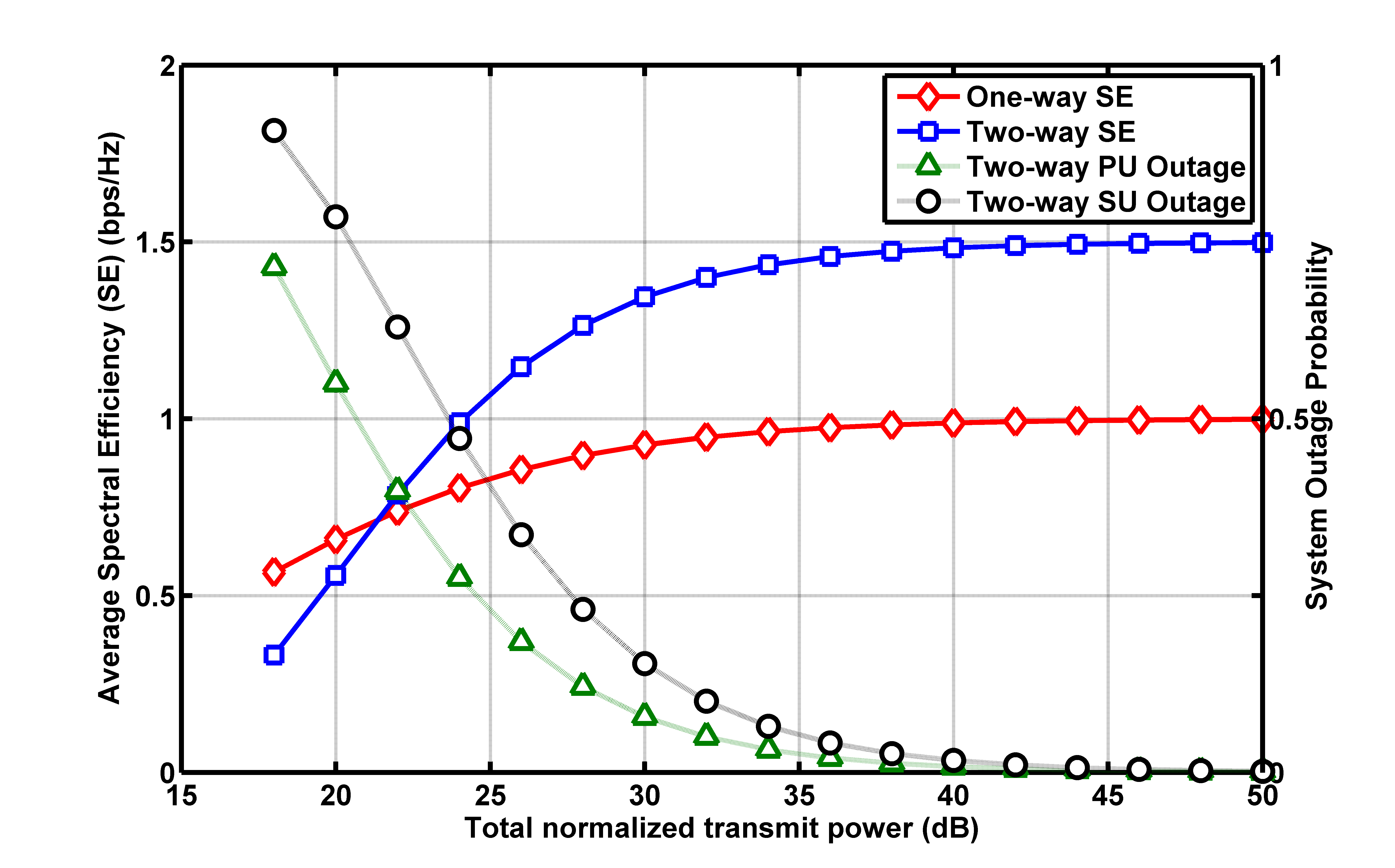}
  \captionof{figure}{Average spectrum efficiency (SE) vs Total normalized transmit power}
  \label{f5}
\end{minipage}%
\end{figure*} 

\tab Fig. \ref{f5} shows the variation of the average SE of the system, following (\ref{e24}), with the normalized transmit power of each PU. It can be clearly seen that SE increases at first and then stabilizes at around $40$ dB transmit power. To explain this nature, the outage variation of the two-way protocol for both PU and SU are also shown in the same figure. It is observed that initially, the outage falls fast as the transmit power increases but then falls slowly beyond $40$ dB. Consequently, SE is found to saturate at around $1.5$ bps/Hz, as is evident from eqn. (\ref{e24}). Moreover, it is also compared with the SE performance of a similar SWIPT-enabled one way relaying (OWR) protocol in \cite{jain2015energy}. To ensure a fair comparison between the two protocols, we consider that the total transmit power of the two protocols as same. Under this assumption, the SE of the two-way protocol is found to be better compared to the one-way protocol, above a transmit power of $21.5$ dB. About $50\%$ higher SE value is achieved in case of two-way protocol, for a total transmit power of around $40$ dB.

\begin{figure*}
\centering
\begin{minipage}{.5\textwidth}
  \centering
  \includegraphics[width=\textwidth]{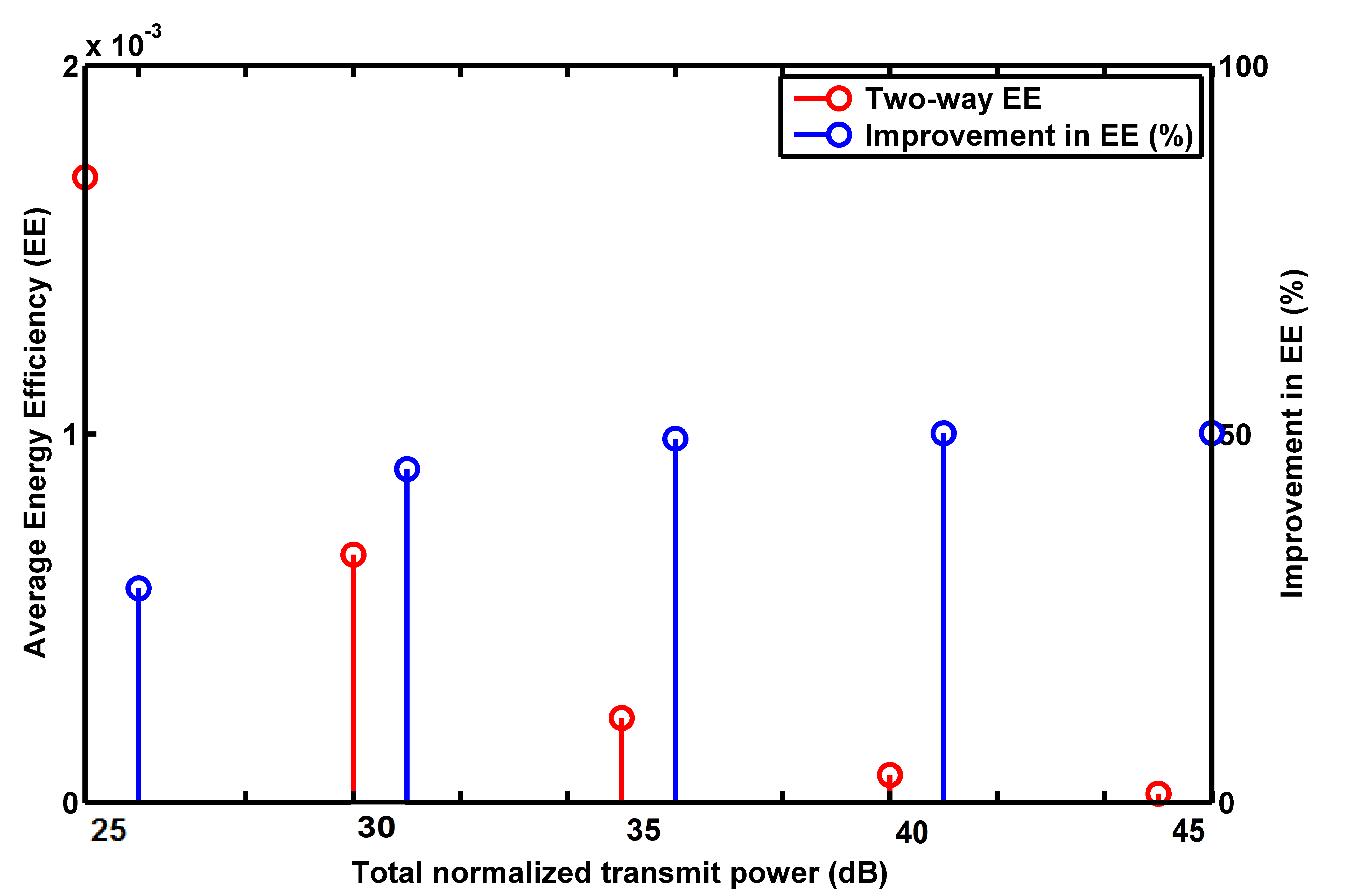}
  \captionof{figure}{Average energy efficiency (EE) for each Total normalized transmit power value}
  \label{f6}
\end{minipage}
\end{figure*}

\tab Fig. \ref{f6} illustrates the average EE of the two-way CR system, following (\ref{e25}), and the corresponding improvement over OWR in \cite{jain2015energy}. As shown in the figure, higher EE is achieved for the two-way protocol with a total normalized transmit power beyond $21.5$ dB. Although EE of the proposed TWR system deteriorates with transmit power, its improvement over the OWR increases with transmit power and is found to be around $50\%$, at a transmit power of 45 dB.\\

\section{Conclusion}
\tab In this paper, we have studied the impact of a PSR protocol on an RF energy harvesting based two-way DF relay network. Analytical expressions for outage probabilities at the destinations are derived and verified using simulations. It is shown that in terms of spectrum and energy efficiency, two-way energy harvesting DF-relay protocol is found to outperform the corresponding one-way protocol. In future, we intend to investigate similar systems with multiple antenna. \\

\ifCLASSOPTIONcaptionsoff
  \newpage
\fi
\vspace{1cm}
\section*{APPENDIX A}
\section*{PROOF OF LEMMA 1 IN (21)}
\vspace{0.5cm}
\hspace{-3mm}This appendix derives $P_{out}^{PU}$, in (21), at the primary users. From eqn. (5), we have
\begin{equation}\tag{A.1.1} \label{a1}
\begin{split}
Pr\big\lbrace R_{SU_1}^{(1)}>R_p\big\rbrace &= Pr\bigg\lbrace \frac{(1-\rho_1) P_{p_{1}}}{d_1^m \sigma^2}\vert h_1 \vert ^2 > 2^{4 R_p}-1 \bigg\rbrace\nonumber\\
&= Pr\bigg\lbrace\vert h_1 \vert ^2 > \frac{\gamma_{p_1} d_1^m \sigma^2}{(1-\rho_1) P_{p_{1}}} \bigg\rbrace\nonumber\\
&= Pr\bigg\lbrace X_1 > \frac{\gamma_{p_1} d_1^m \sigma^2}{(1-\rho_1) P_{p_{1}}} \bigg\rbrace\nonumber\\
&= \int_{\frac{\gamma_{p_1} d_1^m \sigma^2}{(1-\rho_1) P_{p_{1}}}}^{\infty} f_{X_1}(x_1) dx_1\nonumber\\
&= \lambda_1 \int_{\frac{\gamma_{p_1} d_1^m \sigma^2}{(1-\rho_1) P_{p_{1}}}}^{\infty} \text{exp}(-\lambda_1 x_1) dx_1\nonumber\\
&= \text{exp}\bigg\lbrace{-{\frac{\lambda_1 \gamma_{p_1} d_1^m \sigma^2}{(1-\rho_1) P_{p_{1}}}}}\bigg\rbrace
\end{split}
\vspace*{-3mm}
\end{equation}
Similarly, from eqn. (10), we obtain
\begin{equation}\tag{A.1.2} \label{a2}
\begin{split}
Pr\big\lbrace R_{SU_1}^{(2)}>R_p\big\rbrace &= Pr\bigg\lbrace X_2 > \frac{\gamma_{p_1} d_2^m \sigma^2}{(1-\rho_2) P_{p_{2}}} \bigg\rbrace\nonumber\\
&= \text{exp}\bigg\lbrace{-{\frac{\lambda_2 \gamma_{p_1} d_2^m \sigma^2}{(1-\rho_2) P_{p_{2}}}}}\bigg\rbrace .
\end{split}
\end{equation}
where $X_1=\vert h_1 \vert ^2$, $X_2= \vert h_2 \vert ^2$ and \(\gamma_{p_1}=2^{4 R_p}-1\). 
Again, from eqn. (16), we have
\begin{equation} \tag{A.1.3}\label{a3}
\begin{split}
Pr\big\lbrace R_{PU_1}^{(3)}>R_p\big\rbrace &= Pr\bigg\lbrace \frac{a^\prime \lbrace a\vert h_1 \vert ^2 + b\vert h_2 \vert ^2\rbrace \vert h_6 \vert ^2}{b^\prime \lbrace a\vert h_1 \vert ^2 + b\vert h_2 \vert ^2\rbrace \vert h_6 \vert ^2 + 1} > 2^{2 R_p}-1 \bigg\rbrace \\
&= Pr\bigg\lbrace \frac{a^\prime \lbrace a X_1 + b X_2\rbrace X_6}{b^\prime \lbrace a X_1 + b X_2\rbrace X_6 + 1} > \gamma_{p_2} \bigg\rbrace
\end{split}
\end{equation}
where $X_6=\vert h_6 \vert ^2$, $a \overset{\Delta}{=} \frac{\rho_1 P_{p_{1}}}{d_1^m}$, $b \overset{\Delta}{=} \frac{\rho_2 P_{p_{2}}}{d_2^m}$, $a^\prime \overset{\Delta}{=} \frac{\alpha \eta}{2 d_1^m \sigma_{PU_1}^2}$, $b^\prime \overset{\Delta}{=} \frac{(1-\alpha) \eta}{2 d_1^m \sigma_{PU_1}^2}$ and $\gamma_{p_2}=2^{2 R_p}-1\). 
Setting $Y = a{X_1} + b{X_2}$, we can write the probability density function (PDF) of $Y$ as follows :\\
Let $Y \overset{\Delta}{=} U + V$, where $U \overset{\Delta}{=} a{X_1}$ and $V \overset{\Delta}{=} b{X_2}$. Then, the PDF of $U$ and $V$ are given by, \\
\begin{equation}
\hspace{1cm}f_{U}(u) = \frac{1}{a}f_{X_1}{\Big(\frac{u}{a}\Big)}\nonumber\\
\end{equation}
and
\begin{equation}
\hspace{1cm}f_{V}(v) = \frac{1}{b}f_{X_2}{\Big(\frac{v}{b}\Big)}\nonumber\\
\end{equation}\\
Assuming $a > 0$, $b > 0$ and ${X_1}$ and ${X_2}$ are independent random variables that follow exponential distribution with parameters $\lambda_1$ and $\lambda_2$ respectively, we obtain\\
\begin{equation}
\begin{split}
\hspace{2cm}f_{Y}(y) &= \int_{0}^{y} f_{U}(u)f_{V}(y-u)du\nonumber\\ 
&= \frac{1}{ab} \int_{0}^{y} f_{X_1}\Big(\frac{u}{a}\Big)f_{X_2}\Big(\frac{y-u}{b}\Big)du\nonumber\\ 
&= \frac{\lambda_1\lambda_2}{ab} \int_{0}^{y} e^{-\frac{u\lambda_1}{a}}e^{-\frac{(y-u)\lambda_2}{b}}du\nonumber\\
&= \frac{\lambda_1\lambda_2}{ab}e^{-\frac{\lambda_2}{b}y}\int_{0}^{y}e^{-\Big(\frac{\lambda_1}{a}-\frac{\lambda_2}{b}\Big)u}du\nonumber\\
&= \frac{\lambda_1\lambda_2}{ab}e^{-\frac{\lambda_2}{b}y} \frac{ab}{a\lambda_2-b\lambda_1}\Big[e^{-\Big(\frac{\lambda_1}{a}-\frac{\lambda_2}{b}\Big)y}-1\Big]\nonumber\\
\end{split}
\end{equation}
Hence, we have
\begin{equation}\nonumber \label{an1}
\hspace{1cm}f_{Y}(y) = \frac{\lambda_1\lambda_2}{a\lambda_2-b\lambda_1}\Big[e^{-\frac{\lambda_1}{a}y}-e^{-\frac{\lambda_2}{b}y}\Big]
\end{equation}
Hence, (A.1.3) becomes
\begin{equation} \tag{A.1.4}\label{a3}
\begin{split}
Pr\big\lbrace R_{PU_1}^{(3)}>R_p\big\rbrace &= Pr\bigg\lbrace \frac{a^\prime Y X_6}{b^\prime Y X_6 + 1} > \gamma_{p_2} \bigg\rbrace \\
&= Pr\bigg\lbrace X_6 > \frac{\gamma_{p_2}}{(a^\prime - \gamma_{p_2} b^\prime)Y} \bigg\rbrace
\end{split}
\end{equation}
Defining $k \overset{\Delta}{=} \frac{\gamma_{p_2}}{a^\prime - \gamma_{p_2} b^\prime}$, (A.1.4) can be rewritten as
\begin{equation} \tag{A.1.5}\label{a4}
Pr\big\lbrace R_{PU_1}^{(3)}>R_p\big\rbrace = \begin{cases}
1 - Pr\bigg\lbrace X_6 < \frac{k}{Y} \bigg\rbrace,\hspace*{1cm} \gamma_{p_2} < \frac{\alpha}{1-\alpha}\\
1 - Pr\bigg\lbrace X_6 > \frac{k}{Y} \bigg\rbrace = 0, \hspace*{0.4cm} \text{otherwise} .\\
\end{cases}
\end{equation}
The second equality in (A.1.5) is because of the fact that for $\gamma_{p_2} > \frac{\alpha}{1-\alpha}$, the term $k$ will be negative and the probability of an exponential distribution greater than a negative number is always 1. Solving the first equality, when $k$ is positive, using the product of two random variables, we obtain

\begin{equation}\tag{A.1.6} \label{a5}
\begin{split}
Pr\big\lbrace R_{PU_1}^{(3)}>R_p\big\rbrace 
&= Pr\bigg\lbrace X_6 > \frac{k}{Y} \bigg\rbrace,\hspace*{1.8cm} \\
&= \int_{0}^{\infty}Pr\bigg\lbrace X_6 > \frac{k}{Y} \bigg\rbrace f_{Y}(y)dy\\
&= \int_{0}^{\infty}\int_{\frac{k}{y}}^{\infty}f_{X_6}(x_6) f_{Y}(y)dy\\
&= \int_{0}^{\infty}e^{-\frac{\lambda_6 k}{y}}f_{Y}(y)dy\\
&= \frac{\lambda_1\lambda_2}{a\lambda_2-b\lambda_1}\int_{0}^{\infty}e^{-\frac{\lambda_6 k}{y}}\bigg[e^{-\frac{\lambda_1}{a}y}-e^{-\frac{\lambda_2}{b}y}\bigg]dy \\
&= \frac{2 \lambda_1\lambda_2}{a\lambda_2-b\lambda_1} \Bigg[ \sqrt{\frac{\gamma_{p_2} \lambda_6}{(a^\prime - \gamma_{p_2} b^\prime)\lambda_1}a}\hspace{1mm}\mathcal{K}_1{\Bigg(2 \sqrt{\frac{\gamma_{p_2} \lambda_6 \lambda_1}{a(a^\prime - \gamma_{p_2} b^\prime)}}\Bigg)}\\& \hspace{2cm}- \sqrt{\frac{\gamma_{p_2} \lambda_6}{(a^\prime - \gamma_{p_2} b^\prime)\lambda_2}b}\hspace{1mm}\mathcal{K}_1{\Bigg(2 \sqrt{\frac{\gamma_{p_2} \lambda_6 \lambda_2}{b(a^\prime - \gamma_{p_2} b^\prime)}}\Bigg)}\Bigg] .
\end{split}
\end{equation}
Similar to this, from eqn. (17), we obtain
\begin{equation}\tag{A.1.7} \label{a6}
\begin{split}
Pr\big\lbrace R_{PU_2}^{(3)}>R_p\big\rbrace &= \frac{2 \lambda_1\lambda_2}{a\lambda_2-b\lambda_1} \Bigg[ \sqrt{\frac{\gamma_{p_2} \lambda_7}{(a^{\prime\prime} - \gamma_{p_2} b^{\prime\prime})\lambda_1}a}
\hspace{2mm}\mathcal{K}_1{\Bigg(2 \sqrt{\frac{\gamma_{p_2} \lambda_7 \lambda_1}{a(a^{\prime\prime} - \gamma_{p_2} b^{\prime\prime})}}\Bigg)} \\
& \hspace{2cm}-\sqrt{\frac{\gamma_{p_2} \lambda_7}{(a^{\prime\prime} - \gamma_{p_2} b^{\prime\prime})\lambda_2}b}\hspace{2mm}\mathcal{K}_1{\Bigg(2 \sqrt{\frac{\gamma_{p_2} \lambda_7 \lambda_2}{b(a^{\prime\prime} - \gamma_{p_2} b^{\prime\prime})}}\Bigg)}\Bigg] .
\end{split}
\end{equation}
where $X_7=\vert h_7 \vert ^2$. Substituting (A.1.1), (A.1.2), (A.1.6) and (A.1.7) into eqn. (20), we will have the primary outage probability as in (21).

\section*{APPENDIX B}
\section*{PROOF OF LEMMA 2 IN (23)}
\vspace{0.5cm}
\hspace{-3mm}We have already obtained $Pr\big\lbrace R_{SU_1}^{(1)}>R_p\big\rbrace$ and $Pr\big\lbrace R_{SU_1}^{(2)}>R_p\big\rbrace$ in (A.1.1) and (A.1.2) respectively. From eqn. (6), we may write
\begin{equation}\tag{A.2.1} \label{a7}
Pr\big\lbrace R_{SU_2}^{(1)}>R_p\big\rbrace = Pr\bigg\lbrace \frac{1}{4}\log_2 \Bigg(1+\frac{P_{p_{1}}}{d_3^m \sigma_{SU_2}^2}\vert h_3 \vert ^2 \Bigg)>R_p\bigg\rbrace
\end{equation}
Following the steps shown in (A.1.1), we get
\begin{equation}\tag{A.2.2} \label{a8}
\begin{split}
Pr\big\lbrace R_{SU_2}^{(1)}>R_p\big\rbrace &= \text{exp}\bigg\lbrace{-{\frac{\lambda_3 \gamma_{p_1} d_3^m \sigma_{SU_2}^2}{P_{p_{1}}}}}\bigg\rbrace .
\end{split}
\end{equation}
Similarly, from eqn. (11), we obtain
\begin{equation}\tag{A.2.3} \label{a9}
\begin{split}
Pr\big\lbrace R_{SU_2}^{(2)}>R_p\big\rbrace &= \text{exp}\bigg\lbrace{-{\frac{\lambda_4 \gamma_{p_1} d_4^m \sigma_{SU_2}^2}{P_{p_{2}}}}}\bigg\rbrace .
\end{split}
\end{equation}
From eqn. (18), proceeding as in eqn. (\ref{a5}), we get
\begin{equation}\tag{A.2.4} \label{a10}
\begin{split}
Pr\big\lbrace R_{SU_2}^{(3)}>R_s\big\rbrace &=Pr\bigg\lbrace \frac{1}{2} \log_2 \bigg(1+c \big( a\vert h_1 \vert ^2 + b\vert h_2 \vert ^2 \big) \vert h_5 \vert ^2 \bigg)>R_s\bigg\rbrace\\
&= Pr\bigg\lbrace \frac{1}{2} \log_2 \bigg(1+c Y X_5)>R_s\bigg\rbrace\\
&= Pr\Big\lbrace X_5 > \frac{\gamma_s}{cY} \Big\rbrace\\
&= \int_{0}^{\infty}Pr\bigg\lbrace X_5 > \frac{\gamma_s}{cY} \bigg\rbrace f_{Y}(y)dy\\
&= \int_{0}^{\infty}\int_{\frac{\gamma_s}{cy}}^{\infty}f_{X_5}(x_5) f_{Y}(y)dy\\
&= \int_{0}^{\infty}e^{-\frac{\lambda_5 \gamma_s}{cy}}f_{Y}(y)dy\\
&= \frac{2 \lambda_1\lambda_2}{a\lambda_2-b\lambda_1} \bigg[ \sqrt{\frac{\lambda_5 \gamma_s a}{\lambda_1 c}}\hspace{2mm}\mathcal{K}_1{\bigg(2 \sqrt{\frac{\lambda_5 \lambda_1 \gamma_s}{ac}}\bigg)}\\
&\hspace{2cm}-\sqrt{\frac{\lambda_5 \gamma_s b}{\lambda_2 c}}\hspace{2mm}\mathcal{K}_1{\bigg(2 \sqrt{\frac{\lambda_5 \lambda_2 \gamma_s}{bc}}\bigg)}\bigg]
\end{split}
\end{equation}
where $X_5=\vert h_5 \vert ^2$ and $\gamma_s=2^{2 R_s}-1\). Substituting (A.1.1), (A.1.2), (A.2.2), (A.2.3) and (A.2.4) into eqn. (22), we shall have the secondary outage probability as in (23).

\bibliographystyle{IEEEtran}
\bibliography{IEEEabrv,Ref}
\end{document}